# Deposition of Reduced Graphene Oxide Thin Film by Spray Pyrolysis Method for Perovskite Solar Cell

*Manoj Pandey, Dipendra Hamal, Deepak Subedi, Bijaya Basnet,*

*Rajaram Sah, Santosh K.Tiwari, and Bhim Kafle*

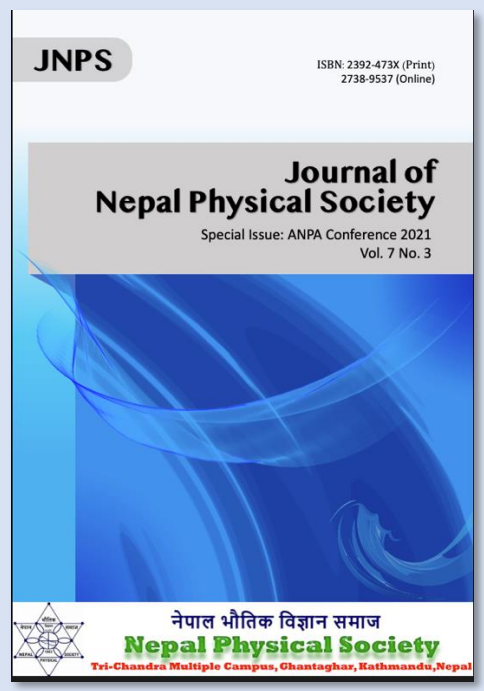

# Deposition of Reduced Graphene Oxide Thin Film by Spray Pyrolysis Method for Perovskite Solar Cell

Manoj Pandey[1], Dipendra Hamal[1], Deepak Subedi[1], Bijaya Basnet[2], Rajaram Sah[3],

Santosh K.Tiwari[4], Bhim Kafle[3]*

[1]*Department of Physics, School of Science, Kathmandu University, Dhulikhel, Nepal*  
[2]*Advanced Materials and Liquid Crystal Institute, Kent State University, Kent, USA*  
[3]*Department of, Chemical Science & Engineering, School of Engineering, Kathmandu University, Nepal*  
[4]*Department of Chemistry, University of Warsaw, Warsaw, Poland*  
***E-mail:** bhim@ku.edu.np*

**Abstract.** The Perovskite absorber layer, the electron transport layer (ETL), the hole transport layer (HTL), and the transparent conducting oxide layer (TCO) are the major components that make up a Perovskite solar cell. Between ETL and HTL, the absorber layer is sandwiched, on which electron-hole pairs are created after absorption of solar radiation. Despite substantial progress toward efficiency, long-term stability still remains a serious concern. Present work focuses toward contributing on the later issue by adopting Titanium dioxide ($TiO_2$) as ETL and reduced graphene oxide (rGO) as HTL. Specifically, in the present work, we report our efforts on the preparation of compact titanium dioxide (C-$TiO_2$) and mesoporous titanium dioxide (M-$TiO_2$) layers as an ETL and a reduced graphene oxide thin film as a HTL. The C-$TiO_2$ film was spin casted on FTO glass followed by casting of M-$TiO_2$ film using Doctor Blading technique. Similarly, the rGO film was produced by spray casting over the glass substrate. The as-prepared ETL and HTL layers were characterized by measuring their optical properties (transmittance and reflectance of thin films). Then, the bandgap, $E_g$ was extracted from reflectance and transmittance curves for ETL and HTL respectively. In the case of rGO, we found the value of $E_g$ to be 2.1 eV, which varies between 2.7eV and 0.02eV depending upon its reduction level based on the previously reported values. Similarly, the bandgap of the C-$TiO_2$ was 4.51 eV which was reduced to 4.12 eV after the addition of M-$TiO_2$, which are 0.9 to 1.1 eV higher than previously reported values. However, bandgap shows decreasing trend after employing M-$TiO_2$ over C-$TiO_2$. In a Perovskite solar cell, both ETL and HTL will be investigated.

## INTRODUCTION

Perovskite solar cell consists of a photosensitizer; methyl ammonium lead halide ($CH_3NH_3PbX_3$; X which is a halogen atom like chlorine, bromine, or iodine). It is thought to be an excellent photo sensitizing material in a solar cell since it absorbs well in the UV – Visible range [1]. This layer is sandwiched between an n-type and p-type material that have an anode and cathode layer, respectively, in a solar cell. When light strikes it, it is absorbed, forming an exciton (electron – hole) pair. As the two carriers move across the ETL/perovskite or HTL/perovskite layer, they dissociate, resulting in a photovoltage [2].

Electron transport is the layer between the active layer and the TCO or cathode in a basic solar cell that prevents carriers (electrons) from recombining with their counterparts in defects in the interface. It must also pass over the UV - visible spectrum without absorption, requiring a large band gap (For example, greater than 3.0eV is required to pass the light with wavelength > 400nm with least absorption) [3]. Hole transport is a layer between the active layer and the anode that prevents carriers (holes) from recombining with their counterparts in interface defects. We can generate a big open circuit voltage in the photo sensitizer and anode junction by introducing an intervening HTL [3-4]. Because of the excellent properties such as high electron mobility of 200,000 $cm^2V^{-1}s^{-1}$, high mechanical strength (200 times greater than the strongest steel), higher transparency (97.7% in the visible region), and flexibility [5-9], graphene, a two-dimensional sheet of carbon atoms arranged in a honeycomb structure, has emerged as a revolutionary material. These characteristics suggest that this material might be used for numerous applications, including energy harvesting and storage [10-22].

Perovskite solar cell exceeds DSSC (dye-sensitized solar cell) in terms of maximum efficiency but not cell stability [23]. A.L Palma et.al. revealed that employing rGO as a hole transporting layer increased the stability of a Perovskite solar cell [24].

## EXPERIMENTAL SECTION

To prepare rGO thin film, we started by mixing 1g of rGO [25] powder in 100 ml solvent which is mixture of 2-propanol and ethanol in the ratio 3:1. The solution is then ultrasonicated for 4 hours and the supernatant of that solution is coated on glass substrate for preparation of thin films by the process of spray pyrolysis method. The deposition is done at the temperature of 80°C for 10 minutes and the height of nozzle is kept 4 cm from the glass substrate. Then the optical analysis of obtained thin film was done using TF Probe Profilometer.

For the preparation of titanium dioxide composite thin film, initially C-TiO$_2$ is spin coated over FTO coated glass and annealed at 450°C for two hours which is followed by deposition of M-TiO$_2$ ( both purchased from Sigma Aldrich) paste by doctor blading method which is again annealed at 450°C for two hours. The optical measurement of obtained films was done by TF Probe Profilometer.

## RESULTS AND DISCUSSION

### I.   Optical measurement of rGO thin film

Figure 1(a) shows the UV-visible transmission spectrum as measured by the profilometer. Swanepoel's model is widely used to determine the optical properties of a thin film from the interference pattern that appears in the transmission spectra [26]. It assumes a very practical case where a thin film is formed over a transparent thick substrate. It also counts all multiple reflections at three interfaces (air-film, film-substrate, and substrate-air) and has been used to compute the optical characteristics of a number of transparent conducting oxides in recent years [27-31], including ZnO, Cu$_3$N, SnO$_2$, titanium oxide, As$_2$Se$_3$, and others. To apply Swanepoel's model, we took the extremes of the interference fringes and used cubic spline interpolation to create two envelopes, $T_M$ and $T_m$, which give the maximum and minimum values of transmission as a continuous function of wavelength, as shown in Fig 1(a).

According to the Swanepoel model [26-27], the refractive index '$n$' of the thin film in the region of the weak and medium absorption (350nm to 1000nm for our rGO film) is,

$$n = \left[ N + \left(N^2 - s^2\right)^{1/2} \right]^{1/2} \qquad \ldots (1)$$

where

$$N = 2s\frac{T_M - T_m}{T_M T_m} + \frac{s^2 + 1}{2}, \quad s = \text{substrate refractive index (1.51 for glass)}$$

Figure 1(b) shows the refractive index of rGO. It illustrates that the refractive index and dielectric constant (square of refractive index) both rise with wavelength. The increasing trend of refractive index with wavelength in the weak absorption region (500nm to 1000nm for our rGO film), where the variation of absorption coefficient is almost negligible (later discussed in Fig 1(c)), tells that total reflection from the thin film to air increases with a wavelength, which automatically decreases the total transmitted light, thus we can see a roughly decreasing transmission with wavelength if we look at the interference-free transmission (which can be termed the average of $T_M$ and $T_m$). As a result, the decreasing transmission pattern with wavelength is explained by an increase in refractive index from 1.5 to 2.0 while traveling from 400nm to 1000nm. In the weak and medium absorption areas, Veronika et al. [32] showed a rising tendency in the refractive index with wavelength for all chemically, thermally, and UV reduced graphene oxide, but they found the reverse trend in graphene oxide. Furthermore, the refractive index values in our investigation ranged from 1.5 to 2.0, which is similar with the results published by Veronika et al.

We used the two neighboring maxima (or minima) at wavelengths $\lambda_1$ and $\lambda_2$, where the refractive indices are $n_1$ and $n_2$, respectively, to calculate the thin film thickness '$d$'. After that, we computed thickness using Equation (2) [26-32] and averaged the thickness values derived from many pairs of two neighboring peaks (maxima) and valleys (minima). 2500nm is the average thickness.

$$d = \frac{\lambda_1 \lambda_2}{2(\lambda_1 n_2 - \lambda_2 n_1)} \quad \ldots (2)$$

Again, from the Swanepoel model [26] in the region of the weak and medium absorption (350nm to 1000nm for our rGO film), the absorption coefficient '$\alpha$' of the thin film is,

$$\alpha = -\frac{\ln(x)}{d} \quad \ldots (3)$$

where

$$x = \frac{F - \left[F^2 - (n^2-1)^3(n^2-s^4)\right]^{1/2}}{(n-1)^3(n-s^2)}$$

$$F = \frac{8n^2 s}{T_i}, \qquad T_i = \frac{2T_M T_m}{T_M + T_m}$$

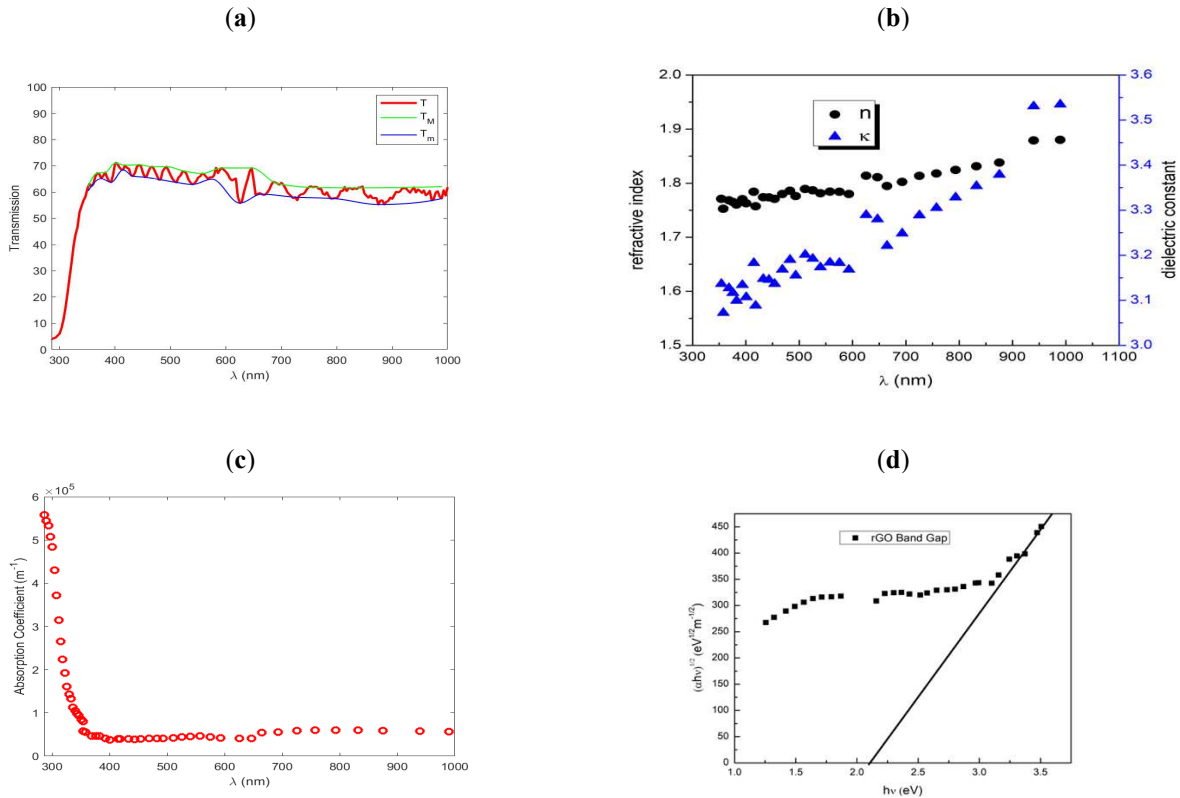

**Figure 1** (**a**) Transmission vs wavelength (**b**) Refractive index and dielectric constant vs wavelength (**c**) Absorption Coefficient vs. wavelength (**d**) Tauc plot of rGO

Figure 1(c) depicts the rGO absorption coefficient. It demonstrates that the absorption coefficient is fairly constant in the weak absorption region (from 450nm to 1000nm), but rapidly increases in the medium absorption region (350nm to 450nm) as one moves closer to the strong absorption region (from 450nm to 300nm). Because of its decreased absorbance between 350nm and 1000nm, reduced graphene oxide film can be a good HTL for Perovskite solar cells.

The Tauc plot [33] is used to determine the optical bandgap ($E_g$), which is represented as

$$\alpha h\nu = A\left(h\nu - E_g\right)^r \qquad \ldots (4)$$

Where,

$h\nu$ is the photon energy, A is a constant, the values of exponent 'r' for direct and indirect transitions are respectively ½ and 2, the value of r for our rGO is ½. The extrapolated line shown at Tauc plot in Figure 1(d) meets the x-axis (i.e. photon energy) at 2.1 eV, indicating that the rGO bandgap is 2.1 eV. According to Velasco et al. [34] and Shen et al. [35], the bandgap of reduced graphene oxide ranges between 2.7eV and 0.02eV depending on the reduction environment (for example, pH) and the degree of reduction (for example time used). Therefore, our rGO's band gap agrees with Velasco et al.

## II. Optical measurement of Titanium dioxide films

Figure 2(a) and (b) shows the reflectance and transmittance sprectrum of C-TiO$_2$ coated over FTO glass and M-TiO$_2$+C-TiO$_2$ coated over FTO glass respectively. Using the Kubelka Munk Function [36], we calculated the band gap energy of C-TiO$_2$ coated over FTO glass and M-TiO$_2$+C-TiO$_2$ coated over FTO glass, and found that the band gap is reduced from 4.51 to 4.12 eV after the addition of M-TiO$_2$ layer as shown in figure 2(c).

According to C. Kahattha et al., M-TiO$_2$ has a bandgap of 3.0-3.2 eV, while it can be reduced by doping non-metallic impurities such as C,N, S, and others over M-TiO$_2$, which can widen the valence band and narrow the bandgap, allowing the surface to generate highly active electrons and holes [37]. The bandgap of FTO ranges between 3.5 and 4.0 eV depending on the deposition conditions and annealing temperature, according to Niu, Ben et al. [38]. At the same time,

Mengjin Yang reported that the C-TiO$_2$ bandgap is 3.38eV, and that its value falls as Ag doping increases [39].

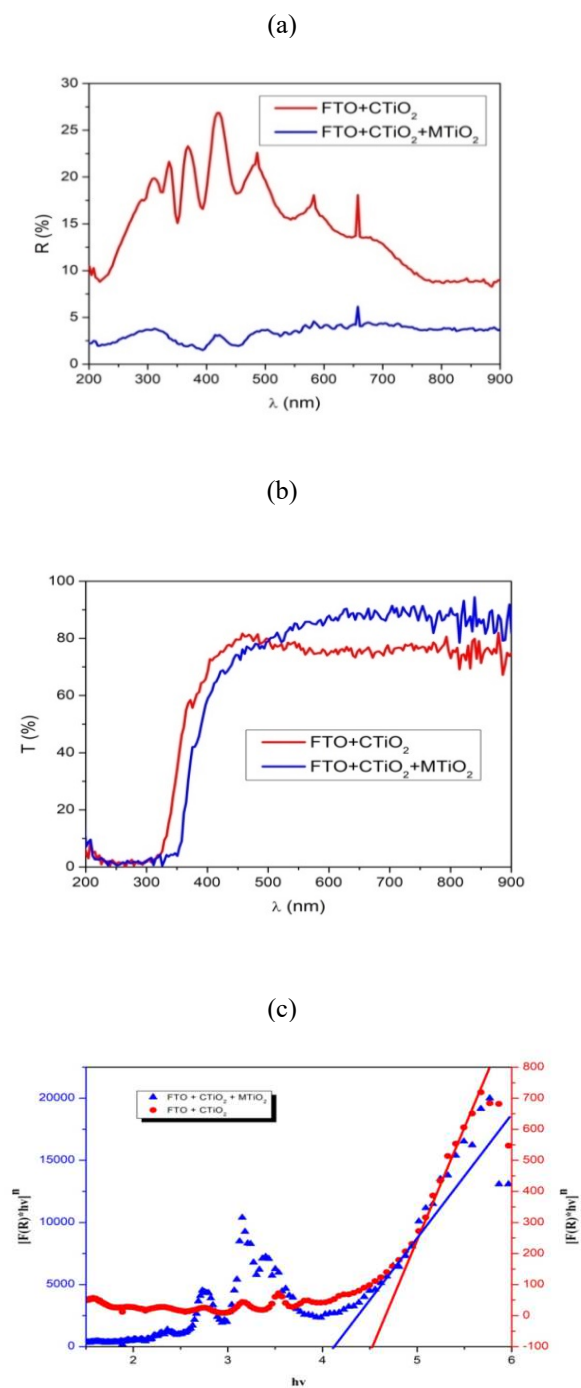

**Figure 2**. (**a**) Reflectance vs Wavelength (**b**) Transmission vs Wavelength (**c**) Kubelka Munk plot indicating C-TiO$_2$ and M-TiO$_2$ bandgap

According to the literatures cited above, C-TiO$_2$ has a larger bandgap than M-TiO$_2$. As a result, a smaller bandgap value is predicted following the addition of the M-TiO$_2$ layer, which we also observed. The bandgap of M-TiO$_2$ and C-TiO$_2$ reported by Kahattha et al. and Mengjin Yang differs from what we observed, which could be either due to the lower TiO$_2$ thickness and higher FTO thickness, or due to the interaction at the FTO and TiO$_2$ interfaces, as bandgap around 4.0 eV is possible for FTO film but not for TiO$_2$ film. On the other hand, because we coated the transparent conducting oxide (TiO$_2$) atop another transparent conducting oxide (FTO) coated glass substrate, a comparison of bandgap values in the same direction may be irrelevant.

## CONCLUSIONS

We have prepared rGO thin film with a novel way of making precursor of rGO from other literatures and found band gap of 2.1 eV and transmittance of approximately 65% in UV-Visible range. This value of band gap is in good agreement with the findings previously reported for its use as HTL in Perovskite solar cell. Whereas, in case of TiO$_2$ ETL, after deposition of M-TiO$_2$ over C-TiO$_2$, the band gap value was reduced from 4.51 eV to 4.12 eV and the transmittance in visible range ($\sim 480 - 800 nm$) increased from approximately 75% to 85%. This indicates that the TiO$_2$ allows major portion of the visible light to the absorber layer in Perovskite Solar cell. Devising complete photocell, employing TiO$_2$ as ETL, Perovskite as absorber and rGO as HTL is in progress.

## AUTHOR CONTRIBUTIONS

Conceptualization: Manoj Pandey and Bhim Kafle; methodology: Manoj Pandey, Rajaram Sah; Software and Laboratory work validation: Manoj Pandey, Bhim Kafle and Bijaya Basnet; formal analysis, Manoj Pandey.; investigation, Bhim Kafle; resources, Bhim Kafle Santosh K. Tiwari ; data curation, Bijaya Basnet.; writing—original draft preparation, Manoj Pandey.; writing—review and editing, Manoj Pandey.; visualization, Manoj Pandey; supervision, Bhim Kafle.; project administration, Deepak Subedi, Dipendra Hamal.; funding acquisition, UGC, Nepal. All authors have read and agreed to the published version of the manuscript.

## ACKNOWLEDGMENTS

We are indebted to Kathmandu University, for providing the access to the laboratory for our research "This research was funded by University Grant Commission, Nepal (UGC Award No.: FRG-76/77-S&T-11).

## CONFLICTS OF INTEREST

"The authors declare no conflict of interest."

## EDITOR'S NOTE

This manuscript was submitted to the Association of Nepali Physicists in America (ANPA) Conference 2021 for publication in the special issue of Journal of Nepal Physical Society.